\documentclass[10pt,twocolumn,twoside]{IEEEtran}
\usepackage{cite}
\usepackage{amsmath,amssymb}
\usepackage{epsfig}
\usepackage{psfrag}
\usepackage{subfigure}
\usepackage[latin1]{inputenc}
\usepackage[T1]{fontenc}
\usepackage{latexsym}
\usepackage{mathrsfs}
\usepackage{graphicx}
\usepackage{booktabs}
\usepackage{color}
\usepackage{enumerate}
\usepackage{multirow}

\newtheorem{proposition}{Proposition}

\begin{document}

\title{Time-Switching Uplink Network-Coded Cooperative Communication with Downlink Energy Transfer}

\author{Guilherme~Luiz~Moritz,~\IEEEmembership{Student~Member,~IEEE,} João~Luiz~Rebelatto,~\IEEEmembership{Member,~IEEE,}
Richard~Demo~Souza,~\IEEEmembership{Senior~Member,~IEEE,}
Bartolomeu~F.~Uchôa-Filho,~\IEEEmembership{Senior~Member,~IEEE,}
and~Yonghui~Li,~\IEEEmembership{Senior~Member,~IEEE}%
\thanks{This work has been supported in part by CNPq, (Brazil).}
\thanks{Guilherme Luiz Moritz, João Luiz Rebelatto and Richard Demo Souza are with the CPGEI, UTFPR, Curitiba, PR, 80230-901, Brazil (e-mail: \{moritz, jlrebelatto, richard\}@utfpr.edu.br).}
\thanks{Bartolomeu~F.~Uchôa-Filho is with the EEL, Federal University of Santa Catarina, Florianópolis, SC, 88040-900, Brazil (e-mail: uchoa@eel.ufsc.br).}%
\thanks{Yonghui Li is with the School of EIE, University of Sydney, Sydney, NSW, 2006, Australia (e-mail: lyh@ee.usyd.edu.au).}
}%

\maketitle

\begin{abstract}
In this work, we consider a multiuser cooperative wireless network where the energy-constrained sources have independent information to transmit to a common destination, which is assumed to be externally powered and responsible for transferring energy wirelessly to the sources. The source nodes may cooperate, under either decode-and-forward or network coding-based protocols. Taking into account the fact that the energy harvested by the source nodes is a function of the fading realization of inter-user channels and user-destination channels, we obtain a closed-form approximation for the system outage probability, as well as an approximation for the optimal energy transfer period that minimizes such outage probability. It is also shown that, even though the achievable diversity order is reduced due to wireless energy transfer process, it is very close to the one achieved for a network without energy constraints. Numerical results are also presented to validate the theoretical results.
\end{abstract}
\begin{keywords}
Cooperative communication, network coding, wireless energy transfer.
\end{keywords}

\section{Introduction} \label{sec:introducao}

Wireless Sensor Networks (WSNs) are generally composed of a large number of small and low-cost sensor nodes, usually operated with small batteries which are difficult, or even impossible, to be replaced or recharged by direct human intervention~\cite{akyildiz.02.wsn}. One promising approach to overcome the limited energy budget of typical WSNs is the energy harvesting technique~\cite{varshney.08,fouladgar12,nasir.13.harvesting,ishibashi.12,zhou.12.harvesting,krikidis.12.harv,zhang.mimo.13,liu.13.et,zhou.12.arxiv}. Among other conventional energy harvesting sources (such as solar power, wind, vibration, etc.), one emerging solution is to harvest energy from the radio-frequency (RF) signals. Although several works have considered ideal receivers that are able to simultaneously harvest energy and process the information received from the same signal~\cite{varshney.08,fouladgar12}, this may be not a practical assumption at the present time. This is because the current technology for RF energy harvesting circuits is not yet able to efficiently harvest energy and decode the information from the same signal~\cite{zhou.12.arxiv,liu.13.et}. In~\cite{zhang.mimo.13}, an approach through time-switching between wireless power and information transfer was proposed, which is ``practically appealing since state-of-the-art wireless information and energy receivers are typically designed to operate separately with very different power sensitivities''. Thus, it is reasonable to consider a receiver with separate energy harvesting and information processing circuits, as in~\cite{ishibashi.12,krikidis.12.harv,zhang.mimo.13,liu.13.et}. Moreover, considering separate information and energy transfer circuits, even operating in different carrier frequencies, allows them to be independently optimized~\cite{mandal.08}.

\begin{figure}[t]
\vspace{-0.4cm}
\centering
{\footnotesize
\subfigure[BP 1: Energy transfer\label{fig:phase1ET}]{
\psfrag{1}[c][]{{\textcolor{white}{S$_1$}}}
\psfrag{2}[][]{{\textcolor{white}{S$_2$}}}
\psfrag{D}[bl][bl][1.2]{{\textcolor{white}{D}}}
\psfrag{a}[c][c]{{\tiny \it \textcolor{red}{Wireless Energy Transfer}}}
\includegraphics[width=0.23\textwidth]{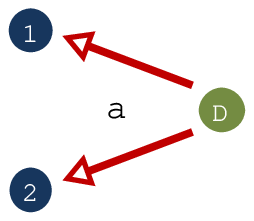}}
\subfigure[BP 2: Broadcast information \label{fig:phase1BC}]{
\psfrag{1}[c][]{{\textcolor{white}{S$_1$}}}
\psfrag{2}[][]{{\textcolor{white}{S$_2$}}}
\psfrag{D}[bl][bl][1.2]{{\textcolor{white}{D}}}
\psfrag{b}[B][B]{$I_1$}
\psfrag{a}[B][B]{$I_2$}
\includegraphics[width=0.23\textwidth]{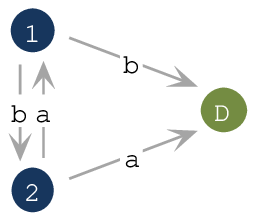}}
\subfigure[CP 1: Energy transfer\label{fig:phase2ET}]{
\psfrag{1}[c][]{{\textcolor{white}{S$_1$}}}
\psfrag{2}[][]{{\textcolor{white}{S$_2$}}}
\psfrag{D}[bl][bl][1.2]{{\textcolor{white}{D}}}
\psfrag{a}[c][c]{{\tiny \it  \textcolor{red}{Wireless Energy Transfer}}}
\includegraphics[width=0.23\textwidth]{fig/harvestAllv4.eps}}
\subfigure[CP 2: Relay information \label{fig:phase2COOP}]{
\psfrag{1}[c][]{{\textcolor{white}{S$_1$}}}
\psfrag{2}[][]{{\textcolor{white}{S$_2$}}}
\psfrag{D}[bl][bl][1.2]{{\textcolor{white}{D}}}
\psfrag{a}[B][B]{$I_1$}
\psfrag{b}[B][B]{$I_2$}
\includegraphics[width=0.23\textwidth]{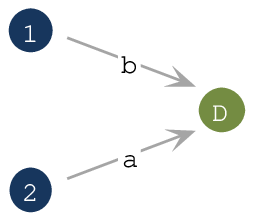}}
}
\caption{Two-source cooperative network with wireless energy transfer. (a,c) The destination charges the sources; (b) the sources broadcast their own information frame (through orthogonal channels) and (d) each user forwards its partner's message to the destination.}
\label{fig:df-harvest}
\vspace{-0.45cm}
\end{figure}


Cooperative communications, a technique initially proposed to increase the diversity order and combat the fading effects inherent to the wireless channel~\cite{sendonaris.03,laneman.04}, has recently been shown to be capable of increasing the energy efficiency of energy-constrained networks ({\it e.g.}~\cite{brante.11}). In cooperative networks, the nodes help each other by relaying their partners' messages. The transmission is usually divided in two phases: the {\it broadcast phase} (BP), where the sources broadcast their own information frames (IFs), and the {\it cooperative phase} (CP), where the nodes transmit parity frames (PFs) to the destination, which are composed of redundant information related to their own IFs and/or to the IFs of their partners. One of the most commonly used cooperative protocols is the decode-and-forward (DF)~\cite{laneman.04}, where the nodes just act as routers in the cooperative phase, relaying the IF from its partner, as illustrated in Fig.~\ref{fig:phase1BC} and Fig.~\ref{fig:phase2COOP}.

Recent works have applied energy harvesting concepts to cooperative wireless networks, generally considering the existence of a dedicated relay~\cite{fouladgar12,nasir.13.harvesting,ishibashi.12}. In~\cite{ishibashi.12}, the authors considered two strategies where the relay could either retransmit the source's information to the destination (while the source harvests the energy of such transmission), or only transfer energy to the source via a RF signal to increase the available source power for the subsequent transmission. In both situations, the source transmission has its reliability improved due to the extra energy harvested/gathered from the relay.

In~\cite{fouladgar12}, under the assumption of perfect channel state information (CSI) at the transmitter, the authors investigated multiuser and multi-hop systems with energy constrained relays for simultaneous wireless information and power transfer. In contrast to~\cite{fouladgar12},~\cite{nasir.13.harvesting} proposed a more practical model by assuming the availability of CSI only at the destination and adopts an architecture with separate information decoding and energy harvesting at the relay. In~\cite{nasir.13.harvesting}, only the amplify-and-forward (AF) relaying was considered, and two protocols were proposed: the time-switching relaying protocol (TSR) and the power-splitting relaying protocol (PSR). Again, only the relay is considered to be energy-constrained, while both source and destination are fully powered. Moreover, it is important to point out that, according to~\cite{ishibashi.twc12}, AF cooperation may impose high peak power levels making DF relaying more practical, especially for energy limited devices.

Another possibility to further increase the performance of a cooperative network is to explore network coding~\cite{ahlswede.00}, by enabling the nodes to transmit linear combinations of two or more IFs in the cooperative phase. Recent works have shown that, if the linear combinations are performed over a large enough finite field GF($q$), greater benefits in terms of diversity order~\cite{xiao.10,rebelatto.10.TIT} or even energy efficiency~\cite{rayel.13.spl} can be achieved.

%

Based on the above arguments, in this work we consider a network where multiple energy-constrained source nodes aim to transmit independent information to a common destination, which is considered to be externally powered and responsible for the wireless energy transfer to the source nodes, as depicted in Fig.~\ref{fig:df-harvest} for a two-source network. We consider a time-switching protocol where the wireless energy transfer from the destination to the source nodes is performed during the first fraction $\alpha$ of each time slot (with $0 \leq \alpha \leq 1)$, and that the energy harvested by the sources (which is used in their transmissions) is a function of the fading realization of inter-user and user-destination channels. It is also considered that the users are able to cooperate through a network coding-based protocol for uplink information transmission.


\subsection{Contributions} \label{subsec:controbutions}

The contributions of this work can be summarized as follows:
\begin{itemize}
\item We provide closed form approximations to the outage probability of some well established cooperative schemes in the case of time-switching wireless energy transfer by considering that the energy harvested by the source (and consequently their transmit power) is a random variable;
\item We derive the diversity order of such schemes, and show that the diversity order is reduced in the energy constrained scenario by only a small amount, remaining near to the one of a network without energy-constraints;
\item Assuming that the channel fading in any link of the network is modeled by a random variable that follows the Rayleigh distribution, we resort to a high date-rate approximation in order to obtain a closed-form equation to the optimal value of the time-sharing parameter $\alpha$ that minimizes the outage probability. Such results are not straightforward and can give important insights on the system design. For example, for a given configuration of the network, we show that choosing an arbitrary value of $\alpha=0.5$ (which would probably be the first choice of most system designers, as the time allocated for energy transfer is the same as the time allocated to information transfer) may lead to a result which is more than 10~dB worse than by choosing its optimal value (which is shown to be $\alpha^{*} \approx 0.15$ in this case);
\item A closed-form approximation (whose accuracy increases with the data-rate) for the $\epsilon$-Outage Capacity (maximum achievable rate for a given target outage probability $\epsilon$) is presented, showing the superiority of the proposed scheme in several scenarios when compared to non-cooperative and DF based cooperative transmissions.
\end{itemize}

The rest of this paper is organized as follows. Section~\ref{sec:system_model} presents the system model. Some preliminaries on cooperative communications are presented in Section~\ref{sec:preliminares}. Section~\ref{sec:etnc} introduces and analyzes the network-coded proposed scheme. Section~\ref{sec:numerical_results} presents some numerical results. Finally, Section~\ref{sec:conclusions} concludes the paper.

{\it Notations:} Throughout this work, $+$ refers to summation over the real field, while $\boxplus$ refers to the summation performed over a finite field GF($q$); $P$ stands for power, while $\mathcal{P}_o$ represents outage probability; $(x)^+$ means $\max\{0,x\}$. Bold lower-case letters denote vectors.

\section{System Model} \label{sec:system_model}

Let us first consider two-user cooperative wireless network composed of two sources (referred to as S$_1$ and S$_2$) aiming to transmit independent information to a common destination (D), as depicted in Fig.~\ref{fig:df-harvest}. In this work, for the sake of simplicity, we assume a symmetric scenario where the nodes are equidistant\footnote{It is important to say that the equidistant assumption facilitates the theoretical analysis, allowing us to get important insights on the actual system performance. Moreover, the numerically evaluated outage probability for the case of non-equidistant nodes may not significantly differ from the case of equidistant nodes~\cite{rayel.13.spl}.}. The transmissions are performed in a time-orthogonal fashion between the users, which operate in the half-duplex mode. A time slot is defined as the time period comprising the transmissions of both S$_1$ and S$_2$.

The packet received by node $j$ after a transmission performed by user $i$ at the time slot $k$ can be written as
\begin{equation}
{\bf y}_j[k] = \sqrt{P_i}[k]h_{ij}[k] {\bf x}_i[k] + {\bf n}_{ij}[k],
\end{equation}
where $i,j \in \{1,2,d\}$ refer to S$_1$, S$_2$ and D, respectively, ${\bf x}_i[k] \in \mathbb{C}^{1 \times N}$ is the transmitted packet of length $N$, $P_i[k]$ corresponds to the transmission power, $h_{ij}[k] \in \mathbb{C}$ represents the block-fading coefficient, whose envelope is modeled as a Rayleigh random variable with zero mean and unitary variance, which remains constant within a time slot and changes in an independent identically distributed (i.i.d.) fashion in both time and space. It is also considered that $h_{ij}[k] \neq h_{ji}[k]$. The additive white Gaussian noise is represented by ${\bf n}_{ij}[k] \in \mathbb{C}^{1 \times N}$, which is assumed to have zero mean and variance $\sigma_j^2/2$ per dimension.

We also assume that the receivers have perfect channel state information (CSI), but the transmitters do not have any CSI. Next, we drop the time index $k$ in order to ease the notation.

The instantaneous signal-to-noise ratio (SNR) is defined as
\begin{equation}
\gamma_{ij} =\bar{\gamma}_{ij}|h_{ij}|^2,
\end{equation}
where $\bar{\gamma}_{ij} =  P_i/\sigma_j^2$ corresponds to the average SNR. Assuming unitary bandwidth and Gaussian inputs, an outage event occurs when the mutual information $I_{ij} = \log_2(1 + \gamma_{ij})$ falls below a given target information rate $\bar{R}$. The probability of such an event is called {\it outage probability}. For the case of Rayleigh fading the instantaneous SNR is exponentially distributed, so that the outage probability is~\cite{goldsmith.05}
\begin{equation} \label{eq:outage_dt}
\begin{split}
\mathcal{P}_{o,ij} &\triangleq \Pr\left\{I_{ij} < \bar{R} \right\}  \\
&= \Pr\left\{|h_{ij}|^2 < \frac{2^{\bar{R}}-1}{\bar{\gamma}_{ij}}\right\}  \\
&= 1-\exp\left(-\frac{2^{\bar{R}}-1}{\bar{\gamma}_{ij}}\right).
\end{split}
\end{equation}

Is it worthy emphasizing that, due to the equidistant assumption made in the system model,~\eqref{eq:outage_dt} corresponds to the outage probability of every single link throughout the network, including the interuser channels, which are not assumed to be outage free.

In order to perform a fair comparison between different protocols, we take into account the multiplexing loss inherent to half-duplex cooperative schemes~\cite{laneman.04}. We consider that the target information rate of the generic cooperative  protocol $X$ is given by $\bar{R}_X \triangleq R/R_X$, where $R$ is the attempted information rate in the case of non-cooperative direct transmission, and $R_X$ corresponds to the code rate of the protocol $X$, defined as the ratio between the number of time slots allocated to the transmission of new data and the total number of time slots used by the protocol, with $0 \leq R_X \leq 1$. For the direct transmission, $R_X=R_{\text{DT}} = 1$.

\section{Preliminaries} \label{sec:preliminares}

\subsection{Cooperative Protocols} \label{subsec:cooperative_protocols}

As already mentioned, in the cooperative protocols the transmission is usually divided in two phases: the {\it broadcast phase}, where the sources broadcast their IFs; and the {\it cooperative phase}, where the sources relay to the destination some PFs, containing redundant information from their partners. In what follows we present two cooperative protocols, namely Decode-and-Forward (DF)~\cite{laneman.04} and nonbinary Network-Coding (NC)~\cite{xiao.10,rebelatto.10.TIT} protocols.

\subsubsection{Decode-and-Forward (DF)} \label{subsubsec:df}

In the DF protocol, after broadcasting their own IF, nodes S$_1$ and S$_2$ retransmit their partner's IF in the cooperative phase, as illustrated in Fig.~\ref{fig:phase1BC} and \ref{fig:phase2COOP}. Thus, the code rate of such protocol is given by $R_{\text{DF}} = 1/2$.
Let us focus on the message from S$_1$ (the same result is valid to S$_2$ due to symmetry). The signals received at S$_2$ and D in the broadcast phase (Fig.~\ref{fig:phase1BC}) are given, respectively, by
\begin{subequations}
\begin{equation} \label{eq:y_r_bc}
{\bf y}_{2} = \sqrt{P_{1}}h_{12}{\bf x_1} + {\bf n}_{2};
\end{equation}
\begin{equation} \label{eq:y_d_bc}
{\bf y}_d = \sqrt{P_1}h_{1d}{\bf x_1} + {\bf n}_d.
\end{equation}
\end{subequations}

The outage probability at S$_2$ is then
\begin{equation} \label{eq:outage_12}
\begin{split}
\mathcal{P}_{o,12} &= \Pr\{|h_{12}|^2 < g_{12}\}  \\
&= 1-\exp\left(-g_{12}\right),
\end{split}
\end{equation}
where $g_{12} = (2^{\bar{R}_{\text{DF}}}-1)/\bar{\gamma}_{12}$.
%

We assume that S$_2$ only relays the message of S$_1$ in the cooperative phase when there is no outage in the interuser channel. Upon receiving two copies of the same message, we consider that the destination performs maximal ratio combining (MRC). If S$_2$ could not recover the message from S$_1$, it retransmits its own message in the cooperative phase. Depending on all the outage patterns in the interuser channels, the overall signals received at the destination containing the message from S$_1$ can then be written as~\cite{laneman.04,xiao.10}
\begin{equation}\label{eq:y_df}
\begin{split}
&{\bf y}_d =\\
&\left\{\!\!\!
  \begin{array}{ll}
    \sqrt{P_1}h_{1d}\,{\bf x_1}\!+\!{\bf n}_{1d},\sqrt{P_2'}h_{2d}'\,{\bf x_1}\!+\! {\bf n}_{2d}' & \hbox{if }  \epsilon_{12}\!=\!\epsilon_{21}\!=\!1; \\[2pt]
    \sqrt{P_1}h_{1d}\,{\bf x_1}\!+\! {\bf n}_{1d}, \sqrt{P_1'}h_{1d}'\,{\bf x_1}\!+\! {\bf n}_{1d}',& \multirow{2}{*}{$\hbox{if }  \epsilon_{12}\!=\!1,\epsilon_{21}\!=\!0;$} \\[1pt]
      \qquad \sqrt{P_2'}h_{2d}'\,{\bf x_1}\!+\! {\bf n}_{2d}' & \\[2pt]
   \sqrt{P_1}h_{1d}\,{\bf x_1}\!+\! {\bf n}_{1d} & \hbox{if }  \epsilon_{12}\!=\!0,\epsilon_{21}\!=\!1; \\[2pt]
    \sqrt{P_1}h_{1d}\,{\bf x_1}\!+\! {\bf n}_{1d}, \sqrt{P_1'}h_{1d}'\,{\bf x_1}\!+\! {\bf n}_{1d}' & \hbox{if } \epsilon_{12}\!=\!\epsilon_{21}\!=\!0.
  \end{array}
\right.
\end{split}
\end{equation}

 In~\eqref{eq:y_df}, the superscript $'$ refers to the cooperative phase ($h_{id}$ and $h_{id}'$ are assumed to be independent) and $\epsilon_{ij} \in \{0,1\}$ is a Bernoulli random variable that represents the occurrence or not of an outage in the channel between nodes $i$ and $j$. If $\epsilon_{ij} = 0$ ($\epsilon_{ij} = 1$), it means that the channel is (is not) in outage.
%
%
The mutual information between ${\bf x_1}$ and ${\bf y}_d$ is
\begin{equation} \label{eq:mutual_df}
I_{\text{DF}} =  \log_2\left(1+ \gamma_{d}\right),
\end{equation}
where
\begin{equation} \label{eq:gamma_d_df}
\gamma_{d} = \bar{\gamma}_{1d} |h_{1d}|^2 + (1\!-\!\epsilon_{21})\bar{\gamma}_{1d}' |h_{1d}'|^2 + \epsilon_{12}\bar{\gamma}_{2d}' |h_{2d}'|^2
\end{equation}
is the instantaneous SNR at node D, obtained from~\eqref{eq:y_df}. Consequently, the outage probability of the DF scheme is given by the sum of the outage probabilities of each individual event from (8), weighted by its probability of occurrence, which leads to~\cite{laneman.04,xiao.10}
\begin{equation} \label{eq:outage_df1}
\begin{split}
\mathcal{P}_{o,\text{DF}} &= \Pr\{I_{\text{DF}} < \bar{R}_{\text{DF}}\} \\
&=\sum_{i=0}^1 \sum_{j=0}^1 \frac{\Pr\{I_{\text{DF}} < R/R_{\text{DF}}|\epsilon_{12}\!=\!i,\epsilon_{21}\!=\!j\}}{(\Pr\{\epsilon_{12}\!=\!i\}\Pr\{\epsilon_{21}\!=\!j\})^{-1}}\\
& \approx  1.5 \mathcal{P}_o^2,
\end{split}
\end{equation}
where $\mathcal{P}_o$ is the outage probability of an individual link as obtained in~\eqref{eq:outage_12}.

\subsubsection{Nonbinary Network Coding (NC)}

In a nonbinary network-coded (NC) based cooperative protocol, instead of simply forwarding, during the cooperative phase the nodes are able to transmit linear combinations of all the available IFs. If such linear combinations are performed over a large enough finite field, it is shown in~\cite{xiao.10} that gains in terms of diversity order can be achieved over the DF scheme.

For example, let us consider that the messages transmitted by sources S$_1$ and S$_2$ in the cooperative phase are non-binary combinations given respectively by $I_1 \boxplus I_2$ and $I_1\boxplus2I_2$, as proposed in~\cite{xiao.10}.

Focusing again on the message from S$_1$, there are four possibilities regarding the occurrence of outage or not in the channels between S$_1$ and S$_2$:
 \begin{enumerate}[I)]
\item If none of the interuser channels is in outage ($\epsilon_{12}=\epsilon_{21}=1$), which occurs with probability $\Pr\{\epsilon_{12}=\epsilon_{21}=1\} = (1-\mathcal{P}_o)^2$, we can see that D is able to recover S$_1$'s message from any two out the following four received packets $(I_1, \; I_2, \; I_1 \boxplus I_2,\; I_1\boxplus2I_2)$. The information packet from S$_1$ is not recovered by D only when the direct transmission and at least two out of the three remaining packets cannot be decoded at the destination, which happens with probability~\cite{xiao.10}
\begin{equation} \label{eq:nc-p1}
\begin{split}
&\hspace{-1cm} \Pr\{I_{\text{NC}} < \bar{R}_{\text{NC}} |\epsilon_{12}\!=\!\epsilon_{21}\!=\!1\} =\\
& \qquad = \mathcal{P}_{o}\left[{3 \choose 2}\mathcal{P}_{o}^2(1-\mathcal{P}_{o}) + \mathcal{P}_{o}^3\right] \\
& \qquad = 3\mathcal{P}_{o}^3 - 2\mathcal{P}_{o}^4 \\
& \qquad \approx 3 \mathcal{P}_{o}^3,
\end{split}
\end{equation}
where ${n \choose k}$ stands for the binomial coefficient and the approximation is valid for the high SNR region;

\item When $\epsilon_{12}\!=\!\epsilon_{21}\!=\!0$ (both interuser channels are in outage), which occurs with probability $\Pr\{\epsilon_{12}=\epsilon_{21}=0\} = \mathcal{P}_o^2$, S$_1$ and S$_2$ retransmit their own messages in the cooperative phase. Upon receiving two copies of the same message, we assume that D performs MRC, leading to the following outage probability~\cite{xiao.10}
\begin{equation}\label{eq:nc-p2}
\Pr\{I_{\text{NC}} < \bar{R}_{\text{NC}} |\epsilon_{12}\!=\!\epsilon_{21}\!=\!0\}\approx \frac{\mathcal{P}_{o}^2}{2}.
\end{equation}

\item When $\{\epsilon_{12}\!=\!1,\epsilon_{21}\!=\!0\}$ (only the channel from S$_2$ to S$_1$ is in outage), which occurs with probability $\Pr\{\epsilon_{12}=1,\epsilon_{21}=0\} = (1-\mathcal{P}_o)\mathcal{P}_o$, S$_1$ is not able to add the message from S$_2$ into its linear combination transmitted in the cooperative phase, such that the set of messages received at D is given by $I_1$, $I_2$, $I_1$ and $I_1\boxplus2I_2$. In such a situation, message $I_1$ is in outage at the destination when both the transmissions of $I_1$ and at least one of the remaining $I_2$ and $I_1 \boxplus 2I_2$ are in outage at the D, which occurs with probability:
\begin{equation}\label{eq:nc-p3}
\begin{split}
& \hspace{-1.5cm} \Pr\{I_{\text{NC}} < \bar{R}_{\text{NC}} |\epsilon_{12}\!=\!1,\epsilon_{21}\!=\!0\} = \\
& \quad = \mathcal{P}_{o}^2 \left(2\mathcal{P}_{o}(1-\mathcal{P}_{o})+\mathcal{P}_{o}^2\right)\\
& \quad \approx 2\mathcal{P}_{o}^3.
\end{split}
\end{equation}
\item When $\{\epsilon_{12}\!=\!0,\epsilon_{21}\!=\!1\}$ (only the channel from S$_2$ to S$_1$ is in outage), which occurs with probability $\Pr\{\epsilon_{12}=0,\epsilon_{21}=1\} = \mathcal{P}_o(1-\mathcal{P}_o)$, S$_2$ is not able to add the message from S$_1$ into its linear combination transmitted in the cooperative phase, so the set of messages received at D is given by $I_1$, $I_2$, $I_1\boxplus I_2$ and $I_2$. In this case, the worst case for message $I_1$ being in outage at D (that is, the lowest number of individual links that must be in outage in order to assure that D will not be able to recover $I_1$, representing the most probable event) occurs when the direct transmission and the packet $I_1\boxplus I_2$ are not correctly recovered by D, which occurs with probability:
\begin{equation}\label{eq:nc-p4}
\Pr\{I_{\text{NC}} < \bar{R}_{\text{NC}} |\epsilon_{12}\!=\!0,\epsilon_{21}\!=\!1\} \approx \mathcal{P}_{o}^2.
\end{equation}

\end{enumerate}

The overall outage probability of the 2-user NC scheme is then obtained by adding up the individual outage probabilities of cases I-IV from~\eqref{eq:nc-p1}-\eqref{eq:nc-p4}, weighted by the probability of occurrence of each case, leading to~\cite{xiao.10,rebelatto.10.TIT}
\begin{equation} \label{eq:outage_nc1}
\begin{split}
\mathcal{P}_{o,\text{NC}} &= \Pr\{I_{\text{NC}} < \bar{R}_{\text{NC}}\} \\
&=\sum_{i=0}^1 \sum_{j=0}^1 \frac{\Pr\{I_{\text{NC}} < \bar{R}_{\text{NC}}|\epsilon_{12}\!=\!i,\epsilon_{21}\!=\!j\}}{(\Pr\{\epsilon_{12}\!=\!i\}\Pr\{\epsilon_{21}\!=\!j\})^{-1}} \\
&= \overbrace{(1-\mathcal{P}_o)^2 \ 3\mathcal{P}_o^3}^{\text{\bf I}} \,\,+ \,\, \overbrace{\mathcal{P}_o^2 \ \mathcal{P}_o^2/2}^{\text{\bf II}} \, + \\
&\quad + \, \underbrace{(1-\mathcal{P}_o)\mathcal{P}_o \ 2\mathcal{P}_o^3}_{\text{\bf III}} \,\,+\,\, \underbrace{\mathcal{P}_o(1-\mathcal{P}_o)\ \mathcal{P}_o^2}_{\text{\bf IV}}\\
&\approx 4\mathcal{P}_{o}^3.
\end{split}
\end{equation}

We can see from~\eqref{eq:outage_nc1} that the diversity order of 3 is achieved for the NC scheme, in contrast to the diversity order of 2 obtained by the DF scheme in~\eqref{eq:outage_df1}. Moreover, note that the code rate of the above NC scheme with two users is $R_{\text{NC}} = 1/2$. For a network with $M$ users, it is shown in~\cite{xiao.10} that diversity order $2M-1$ can be achieved, by adopting a code rate $R_{\text{NC}} = 1/M$.

\subsubsection{Generalized Nonbinary Network Coding (GNC)} \label{subsubsec:gnc}

In~\cite{rebelatto.10.TIT}, the scheme in~\cite{xiao.10} was further generalized by considering $M$ cooperative nodes where each source node is able to broadcast a given number  $k_1$ of IFs in the broadcast phase (as opposed to just one IF in~\cite{xiao.10}), as well as to transmit an arbitrary number $k_2$ of PFs during the cooperative phase, composed of linear combinations over GF($q$) of all the available IFs. It is shown that by choosing the generator matrix of a maximum-distance separable (MDS) error-correcting block code as the combining coefficients, the system can achieve the maximum diversity order~\cite{rebelatto.10.TIT}.

In such a situation, the code rate and the outage probability for a $M$-user GNC scheme are given respectively by~\cite{rebelatto.10.TIT}
\begin{subequations} \label{eq:rate_outage_gnc}
\begin{equation}
R_{\text{GNC}} = \frac{k_1}{k_1 + k_2}; \label{eq:rate_gnc}
\end{equation}
\begin{equation}
\mathcal{P}_{o,\text{GNC}} \approx \mu \, \mathcal{P}_o^{M+k_2} \;\;\;\;\; \mbox{(if $k_2 \ge 2$)} \label{eq:outage_gnc},
\end{equation}
\end{subequations}
where $\mu = {{k_1+k_2-1}\choose{k_2}}$ corresponds to the binomial coefficient. From~\eqref{eq:outage_gnc} we can see that the diversity order of the GNC scheme is $M+k_2$.

\section{Time-Switching Cooperation with Energy Transfer} \label{sec:etnc}
\begin{figure*}[!t]
\centering
{\tiny
\subfigure[EDF\label{fig:time_edt}]{
\psfrag{D}[B][b]{D}
\psfrag{1}[B][b]{S$_1$}
\psfrag{2}[B][b]{S$_2$}
\psfrag{t}[B][B]{{\scriptsize time}}
\psfrag{a}[B][B]{\scriptsize $\alpha T$}
\psfrag{b}[B][B]{\scriptsize $\frac{1-\alpha}{2} T$}
\psfrag{h}[B][B]{{\tiny Energy Transfer}}
\psfrag{p}[B][c]{{\scriptsize Broadcast Phase}}
\psfrag{q}[B][c]{{\scriptsize Cooperative Phase}}
\psfrag{d}[B][B]{$I_1$}
\psfrag{e}[B][B]{$I_2$}
\psfrag{f}[B][B]{$I_2$}
\psfrag{g}[B][B]{$I_1$}
\includegraphics[width=0.49\textwidth]{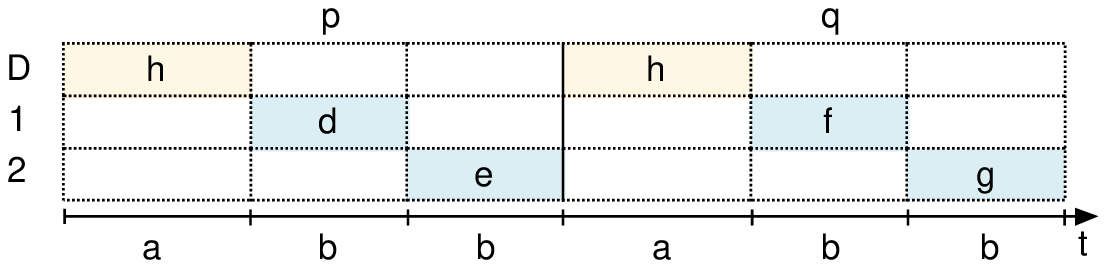}}
\subfigure[ENC\label{fig:time_enc}]{
\psfrag{D}[B][b]{D}
\psfrag{1}[B][b]{S$_1$}
\psfrag{2}[B][b]{S$_2$}
\psfrag{t}[B][B]{{\scriptsize time}}
\psfrag{a}[B][B]{\scriptsize $\alpha T$}
\psfrag{b}[B][B]{\scriptsize $\frac{1-\alpha}{2} T$}
\psfrag{h}[B][B]{{\tiny Energy Transfer}}
\psfrag{p}[B][c]{{\scriptsize Broadcast Phase}}
\psfrag{q}[B][c]{{\scriptsize Cooperative Phase}}
\psfrag{d}[B][B]{$I_1$}
\psfrag{e}[B][B]{$I_2$}
\psfrag{f}[B][B]{$I_1\boxplus I_2$}
\psfrag{g}[B][B]{$I_1\boxplus2I_2$}
\includegraphics[width=0.49\textwidth]{fig/energyTransferTime.eps}}
\subfigure[Two-source EGNC\label{fig:time_egnc}]{
\psfrag{D}[B][b]{D}
\psfrag{1}[B][b]{S$_1$}
\psfrag{2}[B][b]{S$_2$}
\psfrag{t}[B][B]{{\scriptsize time}}
\psfrag{a}[B][B]{\scriptsize $\alpha T$}
\psfrag{b}[B][B]{\scriptsize $\frac{1-\alpha}{2} T$}
\psfrag{h}[B][B]{{\tiny Energy Transfer}}
\psfrag{p}[B][c]{{\scriptsize Broadcast Phase $1$}}
\psfrag{q}[B][c]{{\scriptsize Broadcast Phase $k_1$}}
\psfrag{r}[B][c]{{\scriptsize Cooperative Phase $1$}}
\psfrag{s}[B][c]{{\scriptsize Cooperative Phase $k_2$}}
\psfrag{d}[B][B]{$I_1[1]$}
\psfrag{e}[B][B]{$I_2[1]$}
\psfrag{f}[B][B]{$I_1[k_1]$}
\psfrag{g}[B][B]{$I_2[k_1]$}
\psfrag{x}[B][B]{$\boxplus_1[k_1\!\!+\!\!1]$}
\psfrag{y}[B][B]{$\boxplus_2[k_1\!\!+\!\!1]$}
\psfrag{z}[B][B]{$\boxplus_1[k_1\!\!+\!\!k_2]$}
\psfrag{w}[B][B]{$\boxplus_2[k_1\!\!+\!\!k_2]$}
\includegraphics[width=0.98\textwidth]{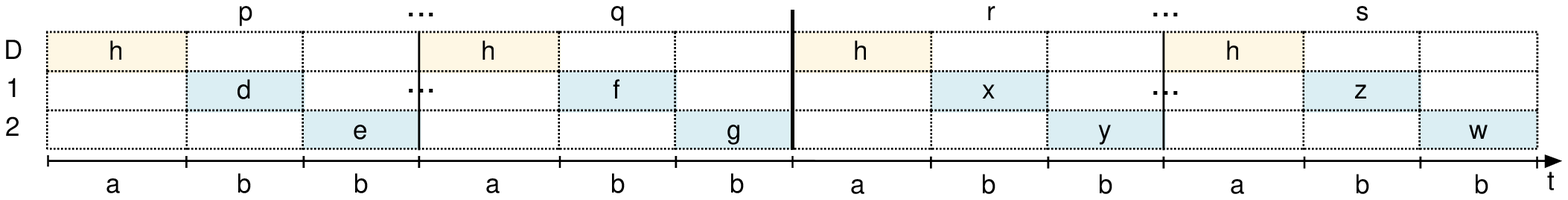}}
}
\vspace{-0.2cm}
\caption{Two-source time division channel allocation with energy transfer considering (a) Energy transfer Decode-and-Forward (EDF) protocol; (b) Energy transfer Network Coding-based (ENC) protocol and (c) Energy transfer Generalized Network Coding-based (EGNC) protocol. In (c), $\boxplus_i[k]$ corresponds to a linear combination transmitted by node $i$ at time slot $k$, which is composed of all the IFs received during the broadcast phase, including node's $i$ own IFs.}
\label{fig:time-division}
\end{figure*}

In the protocols presented in Sec.~\ref{subsec:cooperative_protocols}, it is assumed that the nodes are provided with an unlimited source of energy. However, in some kind of networks, such as wireless sensor networks (WSNs) or wireless body area networks (WBAN), the nodes may be battery powered and experience energy constraints. Since replacing such batteries is not always desirable/possible, this motivates us to investigate the system where the sources do not have internal source of energy in order to perform their transmission. The destination D, in turn, is assumed to be externally powered and capable of transferring energy wirelessly to all the sources~\cite{zhou.12.arxiv,zhou.12.harvesting}. This is illustrated in Fig.~\ref{fig:df-harvest} for a particular case with two-source S$_1$ and S$_2$.

Each time slot is assumed to have a time duration $T$, and $\alpha T$ is the fraction of the time slot in which the destination transfers energy to all the sources, as depicted in Fig.~\ref{fig:time-division} for the particular case of two-source S$_1$ and S$_2$. The remainder of the time slot, $(1-\alpha)T$, is equally divided between all the sources for information transmission~\footnote{In this work, the possibility of any given source harvesting energy during its partner transmission in the cooperative phase is not taken into account. Such energy is neglected since, as commented in~\cite{ishibashi.12}, the power transfer efficiency is maximized for narrowband links that operate at low frequencies while data transmission requires a high data rate with even larger bandwidth, and consequently leads to a lower energy harvesting efficiency.}.


Still focusing on the two-source scenario\footnote{A generalization to $M$-sources will be provided later only for the EGNC scheme, since the DF scheme proposed in~\cite{laneman.04} is limited to a two-source scenario.}, the energy harvested at source node S$_i$ ($i \in \{1,2\}$) during the first portion $\alpha$ of the time slot is
\begin{equation} \label{eq:ehj}
E_i = \eta P_d |h_{di}|^2 \; \alpha T,
\end{equation}
where $\eta$ represents the energy transfer efficiency, with $0 \leq \eta \leq 1$.  Considering that all the energy harvested by node $i$ presented~\eqref{eq:ehj} is consumed during the transmission of the packet ${\bf x}_i$, the available transmit power for S$_i$ is then given by\footnote{
In this work, the power consumed by the circuitry and baseband processing is not taken into account, by assuming that the nodes are provided with a battery capable of keeping their circuitry active. As presented in~\cite{luo.12}, when such power is taken into account and the only source of energy is that harvested by the node, the problem becomes very hard to be modeled.}
\begin{equation} \label{eq:pj}
P_i  = \frac{2 E_i}{(1-\alpha)T} = \frac{2\alpha\eta P_d}{(1-\alpha)} |h_{di}|^2.
\end{equation}

The outage probability with energy transfer in the link between nodes $i$ and $j$ is now defined as
\begin{equation} \label{eq:outage_etdt}
\begin{split}
\mathcal{P}_{o,\text{ET}} &\triangleq \Pr\left\{I_{ij} < R_E \right\}   \\
 &= \Pr\left\{\log_2(1+P_i|h_{ij}|^2) < R_E \right\}  \\
&= \Pr\left\{|h_{di}|^2|h_{ij}|^2 < g\right\} ,
\end{split}
\end{equation}
where $g \triangleq \frac{(1-\alpha)\sigma_d^2(2^{R_E}-1)}{2\alpha \eta P_d}$ and $R_E = \bar{R}_X/(1-\alpha)$ is the adjusted rate threshold such that the overall information rate remains the same when compared to the case of non-cooperative direct transmission without energy transfer. In~\eqref{eq:outage_etdt}, we made use of~\eqref{eq:pj}.

However, from~\eqref{eq:outage_etdt}, we can see that the instantaneous SNR is composed of the product of two random variables with exponential distribution. In~\cite{ahmed.11.mellin}, the authors resort to the Mellin Transform in order to obtain the exact probability density function (pdf) of the product of exponential random variables, resulting in an intricate expression which is hard to manipulate. However, in~\cite{chen.12.approximation} it is shown that the product of exponential random variables can be well approximated by a single random variable with generalized gamma distribution. Thus, according to~\cite{chen.12.approximation}, if $Y_i$ is a random variable with exponential distribution, then the pdf of the random variable $Z = \prod_{i=1}^n Y_i$ can be approximated as
\begin{equation} \label{eq:pdf_gamma1}
f_{Z}(z) \approx \left(\frac{2 m_0}{\Omega_0} \right)^{m_0} \frac{1}{n \Gamma(m_0)} z^{\frac{m_0}{n}-1} e^{-\frac{2m_0}{\Omega_0}z^{1/n}},
\end{equation}
where $m_0 = 0.6102n+0.4263$ and $\Omega_0 = 0.8808n^{-0.9661}+~1.12 $ were heuristically obtained ~\cite{chen.12.approximation}. The cumulative density function (cdf) of $Z$ is given by~\cite{chen.12.approximation}
\begin{equation} \label{eq:cdf_gamma1}
F_{Z}(z) \approx \Gamma\left(m_0,\frac{2m_0}{\Omega_0}z^{1/n}  \right),
\end{equation}
where $\Gamma(a,b) = \frac{1}{\Gamma(a)}\int_0^b e^{-t} t^{a-1}dt$ is the lower incomplete gamma function and $\Gamma(\cdot)$ is the complete gamma function.

The outage probability from~\eqref{eq:outage_etdt} is then re-written by replacing~\eqref{eq:cdf_gamma1} with $n=2$ in~\eqref{eq:outage_etdt}, and given by
\begin{equation} \label{eq:outage_etdt2}
\mathcal{P}_{o,\text{ET}} \approx \Gamma\left(m_0,\frac{2m_0}{\Omega_0}g^{1/2}\right),
\end{equation}
where $m_0 = 1.6467$ and $\Omega_0=1.5709$. For the high SNR region,~\eqref{eq:outage_etdt2} can be well approximated as~\cite[Eq.~(20)]{wang.03.nakagami}
\begin{equation} \label{cdf_gamma1_app}
\mathcal{P}_{o,\text{ET}} \approx \frac{m_0^{m_0-1}}{\Gamma(m_0)}\left(\frac{2}{\Omega_0}g^{1/2}\right)^{m_0}.
\end{equation}

The outage probability of the two-source Energy Transfer Decode-and-Forward (EDF), Energy Transfer Network Coding-based (ENC) and Energy Transfer Generalized Network Coding-based (EGNC) schemes are obtained by replacing~\eqref{eq:outage_etdt2} (or~\eqref{cdf_gamma1_app}) in~\eqref{eq:outage_df1}, \eqref{eq:outage_nc1} and~\eqref{eq:outage_gnc}, respectively, as follows
\begin{subequations} \label{eq:outages_et}
\begin{equation}
\mathcal{P}_{o,\text{EDF}} \approx 1.5\left(\mathcal{P}_{o,\text{ET}}\right)^2; \label{eq:outage_edf}
\end{equation}
\begin{equation}
\mathcal{P}_{o,\text{ENC}} \approx 4\left(\mathcal{P}_{o,\text{ET}}\right)^3; \label{eq:outage_enc}
\end{equation}
\begin{equation}
\mathcal{P}_{o,\text{EGNC}} \approx \mu \left(\mathcal{P}_{o,\text{ET}}\right)^{2+k_2}. \label{eq:outage_egnc}
\end{equation}
\end{subequations}

Note that the outage probabilities in~\eqref{eq:outages_et} can be generally written as $\mathcal{P}_{o,X} \approx \xi_X\left[\mathcal{P}_{o,\text{ET}}\right]^{\mathcal{D}_X}$, where $\xi_X$ and $\mathcal{D}_X$ correspond to the code gain and diversity order of scheme $X$, respectively.


\subsection{Optimal Energy Transfer Time $\alpha^*$}

The outage probability of the EDT, EDF, ENC and EGNC protocols depends on the time-sharing parameter $\alpha$. In what follows, an approximation for the optimal value of $\alpha$ (with respect to the outage probability) is presented.

\begin{proposition}[Optimal Time-Sharing] \label{prop:optimal_alpha}
The value of $\alpha$ that minimizes the outage probability for the scheme $X$ $\in \{$EDT, EDF, ENC and EGNC$\}$ can be approximated as
\begin{equation} \label{eq:optimal_alpha}
\alpha_{X}^* \approx \frac{R_X}{R\ln(2)+R_X}.
\end{equation}
\end{proposition}
\begin{IEEEproof}
See Appendix~\ref{ap:optimal_alpha}.
\end{IEEEproof}

From~\eqref{eq:optimal_alpha}, one can see that $\alpha^*$ does not depend on the SNR, but is an increasing function of $R_X$, as detailed in Appendix~\ref{ap:optimal_alpha}. Intuitively, the increase of $\alpha^*$ with $R_X$ can be explained due to the fact that the outage probability itself increases with $R_X$. Thus, in order to achieve the same outage probability, the larger $R_X$, the larger must be the transmitted power, and consequently the larger the value of $\alpha^*$.

Even though the outage probability varies with $\alpha$, the diversity order does not depend on such time-sharing parameter. In this regard, we present the following proposition.
\begin{proposition}[Diversity Order] \label{prop:diversity_order}
The diversity orders of the two-source EDT, EDF, ENC and EGNC schemes are given respectively by
\begin{subequations} \label{eq:diversity_order_et}
\begin{equation}
\mathcal{D}_{\text{EDT}} = \frac{m_0}{2}, \label{eq:diversity_order_edt}
\end{equation}
\begin{equation}
\mathcal{D}_{\text{EDF}} = m_0, \label{eq:diversity_order_edf}
\end{equation}
\begin{equation}
\mathcal{D}_{\text{ENC}} = \frac{3m_0}{2}, \label{eq:diversity_order_enc}
\end{equation}
\begin{equation}
\mathcal{D}_{\text{EGNC}} = \frac{(2\!+\!k_2)m_0}{2}. \label{eq:diversity_order_egnc}
\end{equation}
\end{subequations}
\end{proposition}
\begin{IEEEproof}
See Appendix~\ref{ap:diversity_order}.
\end{IEEEproof}
From~\eqref{eq:diversity_order_et}, it can be seen that the time sharing between energy transfer and information transmission reduces the diversity order by a factor $m_0/2$ with respect to the case where the sources do not need to be externally powered (which corresponds to a decrease of only $17.7$\%).

\subsection{$\epsilon$-Outage Capacity}

Another performance measure of interest is the $\epsilon$-outage capacity~\cite{goldsmith.05}, defined as the largest achievable rate $\bar{R}_X^{\max}$, for a particular average SNR $\bar{\gamma}$ and optimal time-sharing parameter $\alpha_X^*$, such that the outage probability is less than $\epsilon$. Such a value is presented in the next proposition.
\begin{proposition}[$\epsilon$-outage capacity]\label{prop:maximum_r}
For a particular average SNR $\bar{\gamma}$, the largest achievable rate such that the outage probability is less than $\epsilon$ is given by
\begin{equation} \label{eq:maximum_r}
\bar{R}_X^{\max} = \frac{1}{\ln(2)} \ W\left(\ln(2)\Big[\frac{\epsilon}{\Phi_X}\Big]^{\frac{1}{\mathcal{D}_X}}\bar{\gamma}\right)R_X,
\end{equation}
where $W(\cdot)$ corresponds to the (upper branch) Lambert-W function~\cite{corless.96} (which is an increasing function of its argument), $\mathcal{D}_X$ is the diversity order of the scheme $X$ presented in Proposition~\ref{prop:diversity_order} and $\Phi_X$ is a constant given by
\begin{equation}
\Phi_X = \xi_X \left[\left(\frac{m_0^{m_0-1}}{\Gamma(m_0)}\right)^{\frac{2}{m_0}}\left(\frac{2}{\Omega_0}\right)^{2}\left(\frac{e \ln(2)}{2 \eta}\right)\right]^{\mathcal{D}_X}.
\end{equation}
\end{proposition}
\begin{IEEEproof}
See Appendix~\ref{ap:maximum_r}.
\end{IEEEproof}

The maximum achievable rate in~\eqref{eq:maximum_r} varies logarithmically with the average SNR $\bar{\gamma}$, due to the logarithmic behavior of the Lambert-W function~\cite{corless.96,hassani.05.lambert}. Thus, the code rate $R_X \leq 1$ that multiplies the achievable rate  in~\eqref{eq:maximum_r} has the effect of changing the slope of the curve $\bar{R}_X^{\max} (\bar{\gamma})$. The greater the value of $R_X$, the greater is the increase in the achievable rate provided by an increase in the SNR. In that sense, the DT has the greatest slope, since $R_{\text{DT}}=1$. Thus, it is reasonable to admit the existence of a threshold in the average SNR up to which the cooperative scheme $X \in \{\text{EDF, ENC, EGNC}\}$ outperforms the non-cooperative EDT in terms of achievable rate. In what follows, we present a proposition with an approximation for such a threshold.
\begin{proposition}[Threshold SNR]\label{prop:threshold_snr}
For a particular target outage probability $\epsilon$ and code rate $R_X = \frac{1}{2}$, the cooperative scheme $X$  outperforms the non-cooperative EDT method in terms of achievable rate when the SNR is lower than
\begin{equation} \label{eq:threshold_snr}
\bar{\gamma}_X^{\text{th}} = \frac{1}{\ln(2)} \left(\frac{\epsilon}{\Phi_X}\right)^{-\mathcal{D}_X}\left(\frac{\epsilon}{\Phi_{\text{DT}}}\right)^{-2}.
\end{equation}
\end{proposition}
\begin{IEEEproof}
See Appendix~\ref{ap:threshold_snr}.
\end{IEEEproof}

\subsection{Multi-Source EGNC Scheme}

For the EGNC scheme, the results obtained for the particular case of $M=2$ source nodes are easily extended to a general case of $M>2$ sources by considering that the time allocated to information transmission per user is equal to $(1-\alpha)T/M$. This would lead to a transmission power of source $i$ given by
\begin{equation} \label{eq:pi_m}
P_i  = \frac{M E_i}{(1-\alpha)T} = \frac{M\alpha\eta P_d}{(1-\alpha)} |h_{di}|^2.
\end{equation}

By substituting~\eqref{eq:pi_m} in~\eqref{eq:outage_etdt2} and then replacing the result in~\eqref{eq:outage_gnc}, one can obtain the outage probability of the multi-source EGNC scheme, which is given by:
\begin{equation}\label{eq:outage_m_egnc}
\mathcal{P}_{o,\text{EGNC}} \approx \mu \left[\Gamma\left(m_0,\frac{2m_0}{\Omega_0}g^{1/2}\right)\right]^{M+k_2} \;\;\;\;\; \mbox{(if $k_2 \ge 2$)} ,
\end{equation}
where $g \triangleq \frac{(1-\alpha)\,\sigma_d^2\,(2^{R_E}-1)}{M\,\alpha\,\eta\, P_d}$ and $R_E = \left(\frac{1}{1-\alpha}\right)\left(\frac{k_1 + k_2}{k_1}\right)R$.

Similarly to the two-source case, it can be shown from~\eqref{eq:outage_m_egnc} that the diversity order of the multi-source EGNC scheme is given by:
\begin{equation} \label{eq:diversity_order_egnc}
\mathcal{D}_{\text{EGNC}} = \frac{(M\!+\!k_2)m_0}{2}.
\end{equation}
\section{Numerical results} \label{sec:numerical_results}

In this section, we present some numerical results in order to validate the theoretical results obtained in the previous section. It is considered $\eta=1$ for the energy transfer efficiency\footnote{Although such a value is unrealistic, the conclusions would not change if a different value of $\eta$ is assumed as the effect is equivalent, for all schemes, to changing the average SNR by the same amount.}. The parameters of the EGNC scheme were chosen as $k_1\!=\!k_2\!=\!2$, such that the code rate of all the cooperative schemes was the same.

%

\begin{figure} [!t]
\begin{center}
\includegraphics[width=0.5\textwidth]{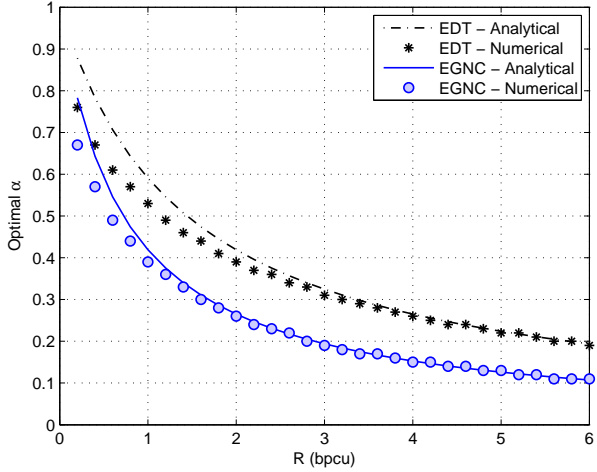}
\end{center}
\caption{Optimal value of $\alpha$ versus the target rate, considering the EDT and EGNC schemes. Note that the optimal $\alpha$ value for the EDF and ENC schemes is the same as the EGNC scheme.}
\label{fig:optimal_alpha}
\end{figure}

Fig.~\ref{fig:optimal_alpha} presents the optimal time-sharing parameter $\alpha^*$ versus the attempted information rate $R$, considering the EDT, EDF, ENC and EGNC schemes. For the ENC and EGNC schemes, the network coding coefficients were chosen from a MDS code according to~\cite{rebelatto.10.TIT}. We can see that $\alpha^*$ decreases as the rate $R$ increases. We can also see that the analytical results match  well with the numerical results, specially when the target rate increases.

\begin{figure} [!t]
\begin{center}
\includegraphics[width=0.5\textwidth]{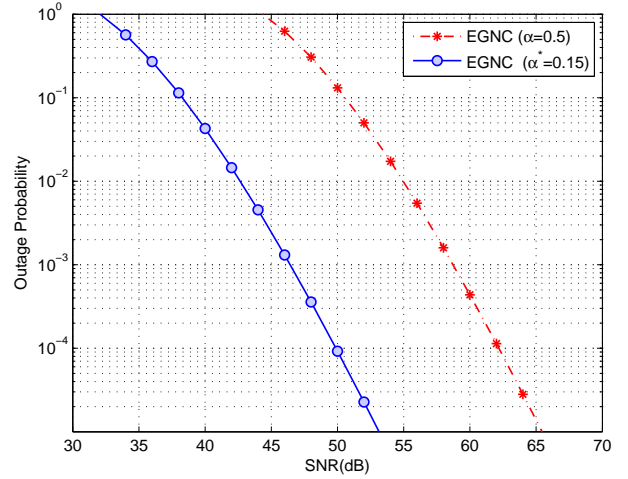}
\end{center}
\caption{Outage probability versus SNR for $R=4$ bps/Hz considering $\alpha=0.5$ and the optimal value $\alpha^{*}=0.15$.}
\label{fig:outage_vs_snr}
\end{figure}

The influence of the time-switching parameter $\alpha$ on the performance of the EGNC scheme is illustrated in Fig.~\ref{fig:outage_vs_snr}, for a scenario with $R=4$ bits per channel use (bpcu). One can see that, for this particular configuration, choosing the optimal value of $\alpha^{*}=0.15$ according to Fig.~\ref{fig:optimal_alpha} can provide a considerable gain in the order of $10$~dB over the option $\alpha=0.5$ (which would be probably the first choice of most system designers, as the time allocated for energy transfer is the same as the time allocated to information transfer). In the next figures, we adopt the optimal value $\alpha^{*}$ for all the schemes.

\begin{figure} [!t]
\begin{center}
\includegraphics[width=0.5\textwidth]{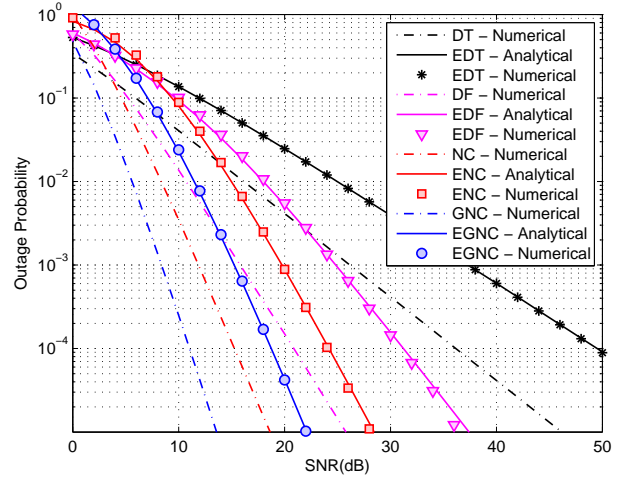}\end{center}
\caption{Outage probability versus SNR considering the EDT, EDF, ENC and EGNC schemes,  for a rate $R=0.5$ bpcu.}
\label{fig:outage_r05}
\end{figure}

\begin{figure} [!t]
\begin{center}
\includegraphics[width=0.5\textwidth]{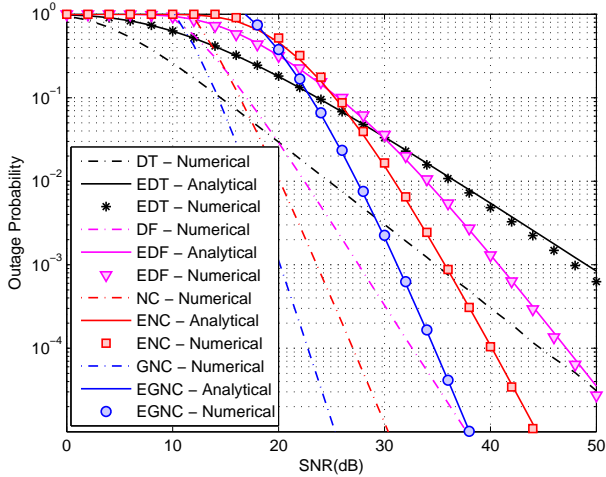}
\end{center}
\caption{Outage probability versus SNR considering the EDT, EDF, ENC and EGNC schemes,  for a rate $R=2$ bpcu.}
\label{fig:outage_r2}
\end{figure}

Fig.~\ref{fig:outage_r05} presents the outage probability versus SNR of the EDT, EDF,  ENC and EGNC schemes, considering a target rate $R=0.5$ bpcu. The outage probability without energy transfer is also presented for comparison. We can see that the cooperative schemes with energy transfer still present  higher diversity orders than the direct transmission without energy transfer. It is also interesting to note that the ENC and EGNC schemes even present higher diversity orders than the DF scheme without energy transfer. One can also see that the analytical results match very well the numerical ones. A result similar to Fig.~\ref{fig:outage_r05} is presented in Fig.~\ref{fig:outage_r2}, but considering a target rate $R=2$ bpcu instead. It can be seen that all the curves are shifted to the right, but the diversity orders are kept unchanged.

\begin{figure} [!t]
\begin{center}
\includegraphics[width=0.5\textwidth]{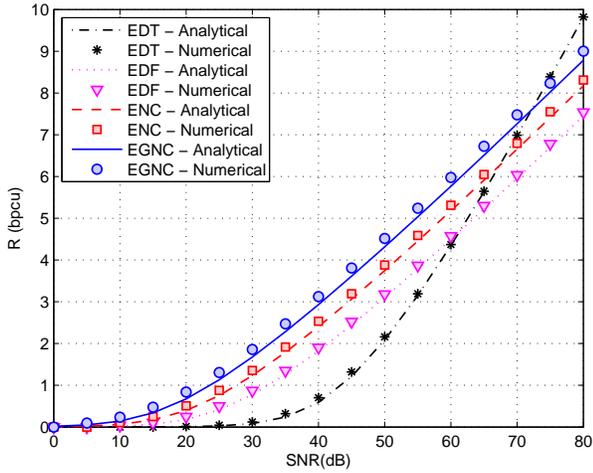}
\end{center}
\caption{Maximum rate versus SNR for the the EDT, EDF, ENC and EGNC schemes, considering a target outage probability $\mathcal{P}_o = 10^{-3}$.}
\label{fig:RvsSNR}
\end{figure}

The maximum achievable rate versus the SNR for the EDT, EDF, ENC and EGNC schemes is presented in Fig.~\ref{fig:RvsSNR}, considering a target outage probability of $10^{-3}$. One can see that the cooperative schemes with energy transfer can achieve higher achievable rates than the non-cooperative EDT for a large SNR range. From Fig.~\ref{fig:RvsSNR} we can also see that, considering the target outage probability as $10^{-3},$ the EDF, ENC, EGNC schemes present higher achievable rates than the EDT when the SNR is lower than $61.9$~dB, $68.4$~dB and $73.8$~dB, respectively. Table~\ref{tab:RvsSNRvsPo} lists the threshold SNR (up to which a given cooperative scheme achieves a higher rate than EDT) for different values of target outage probability. We can see that when the requirement in terms of outage probability is more severe (the target outage probability decreases), the SNR range in which the cooperative protocols outperform the non-cooperative EDT is increased even further.

\begin{table}[!t]
\centering
\caption{\label{tab:RvsSNRvsPo} SNR (in dB) up to which the cooperative schemes outperform EDT in terms of maximum achievable rate, for a given target outage probability $\epsilon$.}
\vspace{-0.2cm}
\begin{minipage}{\columnwidth}
\centering{}
\begin{tabular}{cccc}
\toprule
  $\epsilon$ &  $\bar{\gamma}_{\text{EDF}}^{\text{th}}$ (dB)   &  $\bar{\gamma}_{\text{ENC}}^{\text{th}}$ (dB)   &  $\bar{\gamma}_{\text{EGNC}}^{\text{th}}$ (dB)   \\[3pt]
 &  {\scriptsize Num\,/\,LB~\footnote{Num and LB refer to the numerical values and the lower bound obtained from~\eqref{eq:threshold_snr}, respectively.}}   & {\scriptsize Num\,/\,LB}   & {\scriptsize Num\,/\,LB}   \\
\midrule
  $10^{-3}$ & 61.9\,/\,59.3 & 68.4\,/\,64.0 & 73.8\,/\,68.0  \\
  $10^{-5}$ & 100.7\,/\,95.7 & 110.9\,/\,104.5  & 118.0\,/\,110.5   \\
$10^{-7}$ & 138.7\,/\,132.2 & 152.8\,/\,145.0  & 161.7\,/\,153.0   \\
  \bottomrule
\end{tabular}
\end{minipage}
\end{table}

\begin{figure} [!t]
\begin{center}
\includegraphics[width=0.5\textwidth]{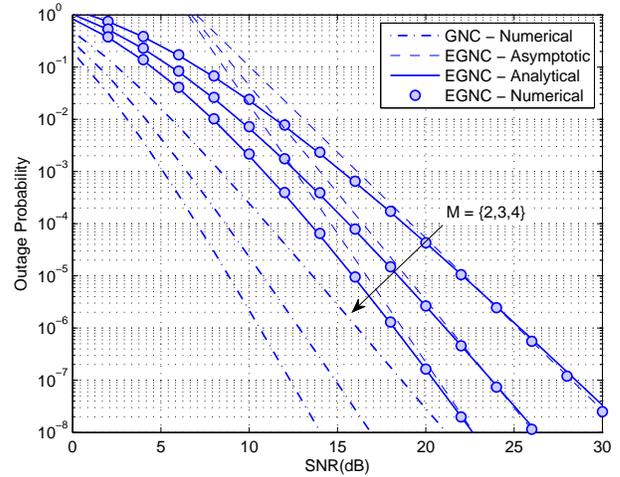}
\end{center}
\caption{Outage probability versus SNR considering the EGNC scheme with $M \in \{2, 3 \text{ and } 4\}$ source nodes, for a rate $R=0.5$ bpcu.}
\label{fig:outageVsSNR_M}
\end{figure}

Fig.~\ref{fig:outageVsSNR_M} presents the outage probability versus the SNR for the EGNC scheme with $M=\{2, 3 \text{ and } 4\}$ source nodes. It can be seen that the diversity order increases when the number of source nodes is increased. In Fig.~\ref{fig:outageVsSNR_M}, ``Asymptotic'' corresponds to the analytical asymptotic behavior of the outage probability with diversity order (slope of the curve) equal to $(M+k_2)m_0/2$ according to~\eqref{eq:diversity_order_egnc}. One can also see that the slope of the numerical and analytical curves support with high precision the claimed diversity order.

\section{Final Comments} \label{sec:conclusions}

We considered a cooperative wireless network where multiple energy-constrained sources aim to transmit independent information to a common destination, which is externally powered and responsible to transfer energy wirelessly to the sources. We obtain a closed-form approximation for the system outage probability of network coding-based cooperative protocols (namely ENC and EGNC), as well as an approximation for the optimal energy transfer time-sharing parameter that minimizes such outage probability. The maximum achievable rate was also obtained analytically, and numerical results demonstrate the accuracy of the analytical derivations. It was shown that, even tough the diversity order is reduced due to the wireless energy transfer, it is less than 20\% lower than a scenario without energy transfer.

\appendices

\section{Proof of Proposition~\ref{prop:optimal_alpha}} \label{ap:optimal_alpha}

Let us first write the outage probability of the scheme $X$ in the general form $\mathcal{P}_{o,X} \approx \xi_X\left[\mathcal{P}_{o,\text{ET}}\right]^{\mathcal{D}_X}$, where $\xi_X$ and $\mathcal{D}_X$ correspond to the code gain and diversity order of scheme $X$, respectively. We then have that:
\begin{equation} \label{eq:outage_appendix}
\begin{split}
\mathcal{P}_{o,X} &\approx \xi_X\left[\mathcal{P}_{o,\text{ET}}\right]^{\mathcal{D}_X}  \\
&= \xi_X\left[\Gamma\left(m_0,\frac{2m_0}{\Omega_0}g^{1/2}\right) \right]^{\mathcal{D}_X} \\
&= \xi_X\left[\Gamma\left(m_0,\frac{2m_0}{\Omega_0}\sqrt{\frac{(1-\alpha)\sigma_d^2\big[2^{\frac{\bar{R}_X}{(1-\alpha)}}-1\big]}{2\alpha \eta_d P_d}}\right) \right]^{\mathcal{D}_X}  \\
&= \xi_X\left[\Gamma\left(m_0,\rho\sqrt{\Lambda_X}\right) \right]^{\mathcal{D}_X},
\end{split}
\end{equation}
where $\rho = \frac{2m_0}{\Omega_0}\sqrt{\frac{\sigma_d^2}{2\eta P_d}}$ and $\Lambda_X=\frac{(1-\alpha)}{\alpha}\Big[2^{\frac{\bar{R}_X}{(1-\alpha)}}-1\Big]$.

Since the lower incomplete gamma function $\Gamma(a,b)$ is a log-convex function in the range $a>0$, $b>0$~\cite[Corollary~3]{qi.02.gamma}, and log-convexity is a stronger condition than function convexity~\cite{andrews.99.functions}, in order to obtain the value of $\alpha$ that minimizes the outage probability in~\eqref{eq:outage_appendix} one must obtain the derivative of~\eqref{eq:outage_appendix} and equate it to zero. By resorting to the fact that $\partial \Gamma(a,b)/\partial b = e^{-b} b^{a-1}$~\cite[Eq.~(8.356.4)]{gradshteyn.07.integrals}, it can be shown that the derivative of $\mathcal{P}_{o,X}$ with respect to $\alpha$ is
\begin{equation} \label{eq:Doutage}
\begin{split}
\frac{\partial \mathcal{P}_{o,X}}{\partial \alpha}
&= \frac{\partial \left[ \xi_X\left[\Gamma\left(m_0,\rho\sqrt{\Lambda_X}\right) \right]^{\mathcal{D}_X}\right]}{\partial \alpha} \\
&= -\frac{\xi_X {\mathcal{D}_X} \rho}{2\sqrt{\Lambda_X}}  \exp\left(-\rho \sqrt{\Lambda_X}\right)\left[\rho\sqrt{\Lambda_X}\right]^{m_0-1}\times\\
&\left[\frac{\Lambda_X - \bar{R}_X \ln(2)2^{\frac{\bar{R}_X}{(1-\alpha)}}}{\alpha(\alpha-1)} \right]\Gamma\left(m_0, \rho \sqrt{\Lambda_X}\right)^{{\mathcal{D}_X}-1}.
\end{split}
\end{equation}

When making~\eqref{eq:Doutage} equal to zero, it reduces to
\begin{equation} \label{eq:Doutage2}
\begin{split}
\Lambda_X^* - \bar{R}_X \ln(2)2^{\frac{\bar{R}_X}{(1-\alpha^*)}} &= 0 \\ \frac{(1-\alpha^*)}{\alpha^*}\Big[2^{\frac{\bar{R}_X}{(1-\alpha^*)}}-1\Big] &= \bar{R}_X \ln(2)2^{\frac{\bar{R}_X}{(1-\alpha^*)}}.
\end{split}
\end{equation}

It is hard to isolate $\alpha^*$ from~\eqref{eq:Doutage2}. However, in order to solve~\eqref{eq:Doutage2}, we can resort to the high-rate approximation
\begin{equation} \label{eq:app_lambda}
\Lambda_X^* \approx \frac{(1-\alpha^*)}{\alpha^*}2^{\frac{\bar{R}_X}{(1-\alpha^*)}}.
\end{equation}

Thus, by applying~\eqref{eq:app_lambda} in~\eqref{eq:Doutage2}, it is easy to show that the optimal value of $\alpha$ becomes
\begin{equation} \label{eq:Doutage3}
\alpha^* \approx \frac{1}{\bar{R}_X \ln(2)+1}.
\end{equation}

Finally,~\eqref{eq:optimal_alpha} is obtained by replacing $\bar{R}_X = R/R_X$ in~\eqref{eq:Doutage3}. 

\section{Proof of Proposition~\ref{prop:diversity_order}} \label{ap:diversity_order}
The diversity order is defined as~\cite{goldsmith.05}
\begin{equation} \label{eq:diversity_order}
\mathcal{D}_X = \lim_{\bar{\gamma}\rightarrow \infty}-\frac{\log \mathcal{P}_{o,X}}{\log \bar{\gamma}},
\end{equation}
where $\mathcal{P}_{o,X}$ corresponds to the overall outage probability of the scheme $X$, and $\bar{\gamma}$ to the average SNR. By replacing the outage probabilities from~\eqref{cdf_gamma1_app} and~\eqref{eq:outages_et} in~\eqref{eq:diversity_order}, it is easy to show that the diversity order of the EDT, EDF, ENC and EGNC schemes are the ones presented in~\eqref{eq:diversity_order_et}. 

\section{Proof of Proposition~\ref{prop:maximum_r}} \label{ap:maximum_r}

We first substitute the optimal value of the time-sharing parameter $\alpha^*$ obtained from~\eqref{eq:optimal_alpha} in the asymptotic approximation of the outage probability from~\eqref{cdf_gamma1_app}, leading to
\begin{equation} \label{eq:outage_appendix2}
\mathcal{P}_{o,\text{ET}} \approx \frac{m_0^{m_0-1}}{\Gamma(m_0)}\left(\frac{2}{\Omega_0}\left[\frac{(e2^{\bar{R}_X}-1)\bar{R}_X}{2 \eta \bar{\gamma}}\right]^{1/2}\right)^{m_0}.
\end{equation}

By writing the outage probabilities from~\eqref{eq:outages_et} in the form $\mathcal{P}_{o,\text{X}} = \xi_X [\mathcal{P}_{o,\text{ET}}]^{\mathcal{D}_X}$ and resorting to the high-rate approximation $e\,2^{\bar{R}_X}-1 \approx e\,2^{\bar{R}_X}$, we have that
\begin{equation} \label{eq:outage_appendix3}
\mathcal{P}_{o,\text{X}} \approx \xi_X \left[ \frac{m_0^{m_0-1}}{\Gamma(m_0)}\left(\frac{2}{\Omega_0}\left[\frac{e\,2^{\bar{R}_X}\bar{R}_X}{2\, \eta \,\bar{\gamma}}\right]^{1/2}\right)^{m_0}\right]^{\mathcal{D}_X}.
\end{equation}

The maximum rate value from~\eqref{eq:maximum_r} is then obtained by setting the target outage probability $\mathcal{P}_{o,\text{X}} = \epsilon$ in \eqref{eq:outage_appendix3} and isolating $\bar{R}_X$. In order to isolate $\bar{R}_X$ from~\eqref{eq:outage_appendix3}, we resort to the (upper branch) Lambert-W function $W(z)$, which is defined as the inverse function of $z e^z$ ~\cite{corless.96}.

\section{Proof of Proposition~\ref{prop:threshold_snr}} \label{ap:threshold_snr}

  We are interested in finding the threshold value $\bar{\gamma}^{\text{th}}$ of the average SNR for which the $\epsilon$-outage capacity for the scheme $X$ equals the maximum achievable rate of the EDT scheme. That is, the value of $\bar{\gamma}$ for which:
\begin{equation} \label{eq:maximum_r_t1}
\bar{R}_X^{\max} - \bar{R}_{\text{DT}}^{\max} = 0.
\end{equation}

Thus, by replacing~\eqref{eq:maximum_r} in~\eqref{eq:maximum_r_t1} and after a few manipulations, one must have that:
\begin{equation} \label{eq:maximum_r_t2}
W\left(\ln(2)\Big[\frac{\epsilon}{\Phi_X}\Big]^{\frac{1}{\mathcal{D}_X}}\bar{\gamma}^{\text{th}}\right)R_X - W\left(\ln(2)\Big[\frac{\epsilon}{\Phi_{\text{DT}}}\Big]\bar{\gamma}^{\text{th}}\right) = 0.
\end{equation}
It is hard (if possible) to isolate $\bar{\gamma}^{\text{th}}$ from~\eqref{eq:maximum_r_t2} without resorting to any approximation.
In~\cite{hassani.05.lambert}, it is shown that, for $z>e$, the Lambert-W function can be upper bounded by
\begin{equation} \label{eq:upper_bound_lambert}
W(z) < \ln(z),
\end{equation}
such that the lower the value of $z$, the tighter is the bound in~\eqref{eq:upper_bound_lambert}~\cite{hassani.05.lambert}. Thus, if we approximate $W(z) \approx \ln(z)$ in~\eqref{eq:maximum_r_t2}, we would have
\begin{equation} \label{eq:maximum_r_t3}
\underbrace{\ln\left(\ln(2)\Big[\frac{\epsilon}{\Phi_X}\Big]^{\frac{1}{\mathcal{D}_X}}\bar{\gamma}^{\text{th}}\right)R_X}_{\text{A}} - \underbrace{\ln\left(\ln(2)\Big[\frac{\epsilon}{\Phi_{\text{DT}}}\Big]\bar{\gamma}^{\text{th}}\right)}_{\text{B}} = 0.
\end{equation}

From the definition of $\bar{\gamma}^{\text{th}}$, one have that $\bar{R}_X^{\max}>\bar{R}_{\text{DT}}^{\max} \; \forall \; \bar{\gamma} < \bar{\gamma}^{\text{th}}$. Since $0<R_X<1$ in~\eqref{eq:maximum_r_t2}, we can say that the argument of the approximation $\ln(\cdot)$ in term A  is larger than the argument of B in~\eqref{eq:maximum_r_t3}, since the $\ln(\cdot)$ is a strictly increasing function. As a result, the tightness of the bound from term A is lower than the tightness of the bound regarding term B.
Thus, the result of the subtraction in~\eqref{eq:maximum_r_t3}, when performed for values of $\bar{\gamma} < \bar{\gamma}^{\text{th}}$, would be strictly larger than the one in~\eqref{eq:maximum_r_t2}, which would lead to an upper bound.

The average SNR $\bar{\gamma}^{\text{th}}$ isolated from~\eqref{eq:maximum_r_t3}, assuming the particular case of the EDF, ENC and EGNC schemes where $R_X = 1/2$, would then lead to the threshold SNR as presented in~\eqref{eq:threshold_snr} which is strictly lower (lower bound) than the one obtained by isolating $\bar{\gamma}^{\text{th}}$ from~\eqref{eq:maximum_r_t2}.

\IEEEtriggeratref{12}
\bibliographystyle{IEEEtran}
\bibliography{IEEEabrv,biblio}

\end{document}